\documentclass[useAMS,usenatbib]{mn2e}
\usepackage{times}
\usepackage{aas_macros}
\usepackage{graphicx}

\def\wig#1{\mathrel{\hbox{\hbox to 0pt{%
          \lower.6ex\hbox{$\sim$}\hss}\raise.4ex\hbox{$#1$}}}}

\def\mstar{M_\star}

\def\msol{\rm M_\odot}

\def\esc{{\rm esc}}

\title[A late formation of Jupiter]{Jupiter's composition: sign of a
  (relatively) late formation in a chemically evolved protosolar disk} 

\author[T. Guillot and R. Hueso]{Tristan Guillot$^{1}$\thanks{E-mail:
    guillot@obs-nice.fr (TG); wubhualr@lg.ehu.es (RH)} and Ricardo
    Hueso$^2$\\
    $^1$ CNRS UMR 6202, Observatoire de la C\^ote d'Azur, BP 4229,
    06304 Nice CEDEX 04, FRANCE\\ 
    $^2$ F\'{\i}sica Aplicada I, E.T.S. Ing., Universidad Pais Vasco,
    Alda. Urquijo s/n. 48013 Bilbao, Spain}
\begin{document}

\date{Submitted to MNRAS: 7 October 2005; Revised: \today}

\pagerange{\pageref{firstpage}--\pageref{lastpage}} \pubyear{2005}

\maketitle

\label{firstpage}

\begin{abstract} 
It has been proposed that the enrichment in noble gases found by
Galileo in Jupiter's atmosphere can be explained by their delivery
inside cold planetesimals. We propose instead that this is a sign that
the planet formed in a chemically evolved disk and that noble gases
were acquired mostly in gaseous form during the planet's envelope
capture phase.

We show that the combined settling of grains to the disk midplane in
the cold outer layers, the condensation of noble gases onto these
grains at temperatures below 20-30K, and the evaporation from high
disk altitudes effectively lead to a progressive, moderate enrichment
of the disk. The fact that noble gases are vaporized from the grains
in the hot inner disk regions (e.g. Jupiter formation region) is not a
concern because a negative temperature gradient prevents convection
from carrying the species into the evaporating region. We show that
the $\sim 2$ times solar enrichment of Ar, Kr, Xe in Jupiter is hence
naturally explained by a continuous growth of the planet governed by
viscous diffusion in the protosolar disk in conjunction with an
evaporation of the disk and its progressive enrichment on a million
years timescale.
\end{abstract}

\begin{keywords}
Solar system: formation, planetary systems, planetary
systems: formation, planetary systems:protoplanetary disks, accretion
disks, planets: Jupiter, planets: Saturn
\end{keywords}

%________________________________________________________________
\section{Introduction}

Ten years have elapsed since the Galileo probe entered Jupiter's
atmosphere and allowed {\it in situ} measurements of its composition
\citep[e.g.,][]{1996Sci...272..837Y, 1999Natur.402..269O}. One of the
key results of the mission is that, when compared to hydrogen, most
measurable species appear to be enriched by a factor two to four
compared to a solar composition.

%% A few exceptions exist and correspond to species which are depleted
%% compared to the solar composition: Helium and neon, which are believed
%% to sink deeper inside the planet due to a phase separation
%% \citep[e.g.,][]{1977ApJS...35..239S, 1995EOS..76..343}, and water,
%% whose low measured abundance is generally taken as a sign of an active
%% meteorology associated to a dry region \citep{1998Icar..132..205S}.

Most surprisingly, the list of enriched species includes not
only ``ices'' (i.e. species that condense at temperature $\sim 100\,$K
in the protosolar disk like methane, ammonia), but also the noble
gases Ar, Kr, Xe \citep{2000JGR...10515061M}, which are enriched
compared to the new solar abundances \citep[e.g.][for a
review]{2003ApJ...591.1220L} by factors $2.2\pm 0.8$, $2.0\pm 0.7$,
$2.0\pm 0.5$, respectively. This led \citet{1999Natur.402..269O} to
propose that Jupiter formed from very low-temperature planetesimals,
at a temperature $\sim 30\,$K allowing a direct condensation of noble
gases onto amorphous ice. This explanation was challenged by
\citet{2001ApJ...550L.227G} who proposed that noble gases would be
incorporated into crystalline ice by clathration at slightly higher
temperatures, and that they would thus be delivered into Jupiter by
icy planetesimals. The large amount of cages available to trap noble
gases implies that the ratio of oxygen to other species in Jupiter
should be highly non-solar, but the precise factor depends on the
condensation sequence \citep{2001ApJ...550L.227G, 2004P&SS...52..623H,
2005ApJ...622L.145A}.

%% Another proposed explanation
%% \citep{2004ApJ...611..587L} is that Jupiter would have formed in a
%% warm, carbon-rich, water-poor environment, and that both hydrogen and
%% helium would have both sunk to the deeper regions of the planet. This
%% last explanation is physically challenging, given that the lighter gas
%% would have to work against gravity.

%% Of course, the abundances measured in Jupiter's atmosphere may not
%% apply to the planet as a whole if there is a first order phase
%% transition between molecular and metallic hydrogen. In that case, the
%% abundances could be modified in a way that will depend on the
%% different chemical potentials at pressures $\sim 1$\,Mbar
%% \citep[e.g.,][]{1989oeps.book..539H}. However, if the effect is
%% significant, there is a priori no reason that it should have the same
%% sign and the same magnitude for species that are so different in terms
%% of abundances and of chemical and electronic structures. On this
%% basis, we will assume that appart from helium and neon, and from refractory
%% species including water, that the measured atmospheric abundances
%% pertain to the entire planet, to the possible exception of the central
%% core \citep[e.g.,][]{2005AREPS..33..493G}.  

In this letter, we propose instead that the present abundances of
Jupiter indicate that the planet (and other giant planets as well)
formed in an chemically evolved protoplanetary disk, in which part of
the light gases (hydrogen, helium and maybe neon) had been lost by
photoevaporation.

Except as otherwise noted, we will use the disk evolution model
described by \citet{HuGu05} with the following values of the
parameters: $M_{\rm cd}=1\,\msol$ (mass in the molecular cloud
core), $M_0=0.4\,\msol$ (seed mass for the protosun), $T_{\rm
cd}=10\,$K (ambient interstellar temperature), $\omega_{\rm
cd}=10^{-14}\,\rm s^{-1}$ (assumed uniform rotation rate of the cloud
core), $\alpha=0.01$ (viscosity).

\section{Solids as carriers of the noble gases?}

\begin{figure}
\hspace*{-0.5cm}\resizebox{9cm}{!}{\includegraphics[angle=0]{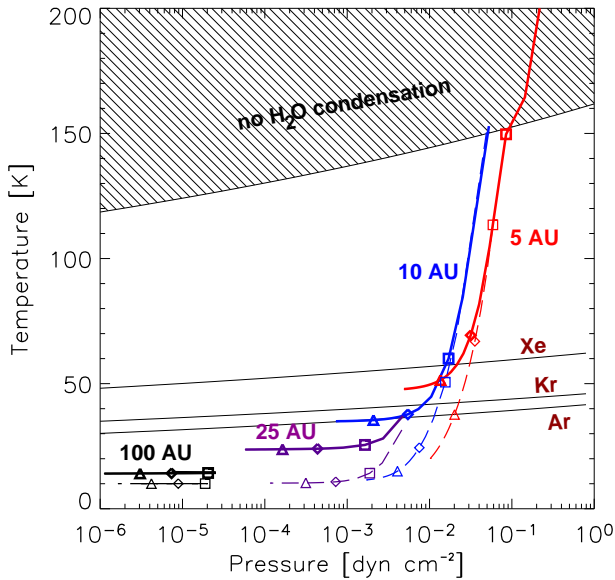}}
\caption{Temperature pressure diagram showing the region of
  condensation of H$_2$O, and the clathration lines for Xe, Kr and Ar
  into crystalline ice, assuming a solar composition. The coloured
  lines are evolution tracks in the protosolar disk at distances of 5,
  10, 25 and 100\,AU, respectively, for irradiated disks (plain)
  or without including stellar irradiation (dashed). The symbols
  corresponds to ages of 1 (squares), 2.5 (diamonds) and 5 Ma
  (triangles), respectively. The disk evolution models are calculated
  without evaporation.}
\label{fig:clathrates}
\end{figure}

Explaining the efficient delivery of noble gases into Jupiter's
atmosphere by embedding them into solids (either through a direct
condensation at $\sim 30$\,K, or through clathration at slightly
higher temperature) is difficult: As shown by
Fig.~\ref{fig:clathrates}, the disk should always remain warmer due to
stellar irradiation up to 10 AU from the central star, i.e. where
Jupiter and Saturn should have formed. A possibility could be that the
disk self-shadows this region \citep[e.g.,][]{2001ApJ...560..957D},
but it remains to be proved that this effect occurs at the right time
and distances, and that it is able to maintain sufficiently low
temperatures. Similarly, Jupiter could have formed further from the
Sun and migrated inward \citep*{2005ApJ...622L.145A}, but it is yet
unclear why Jupiter would have accreted its gas in the $>10$\,AU
region, sufficiently late for the disk to have cooled, and then not in
the $5-10$\,AU region when its migration there imposes that a massive
disk was still present. 

Furthermore, the clathration model requires a large number of
available cages to trap a significant fraction of the argon. This
implies that the ratio of H$_2$O to all other clathrated species has
to be large to provide as many cages as possible, and yet it is
constrained by interior models of Jupiter that imply that the total
mass of heavy elements in Jupiter, including the core and including
rocks, is smaller than $40\rm\,M_\oplus$
\citep{2005AREPS..33..493G}. Alibert et al. explain the observed
abundances of noble gases by the accretion of $\sim 20\rm\,M_\oplus$
of clathrated ices, assuming a 100\% clathration efficiency. An
efficiency of 50\% or less would require accreting a larger amount of
ices, in violation of the interior models constraints. 
%Finally, this overabundance of H$_2$O has to be
%postulated but is not consistently calculated.

In summary, although it appears very likely that chemical species
including noble gases were at some point trapped into solid
planetesimals, explaining Jupiter's composition through that mechanism
only seems to require an unlikely combination of factors.

\section{Giant planet formation in a chemically evolved disk}

We now assess the possibility that the giant planets were formed in a
chemically evolved protoplanetary disk and that noble gases were
delivered to the planets mostly in {\it gaseous} form. In our model,
the noble gases are trapped into the solids at low temperatures in the
outer disk, but most of them are released in gaseous form in the giant
planet formation region. We postulate that this trapping is almost
complete at 30 AU and beyond because of the low ($\sim 20$\,K)
temperatures \citep[see][and
fig.~\ref{fig:clathrates}]{2005Icar..175..546N}.

%% Let us consider a disk of mass $\mdisk(t)$ that is evolving viscously
%% and losing mass by accretion onto the star at a rate $\mdstar$. The
%% disk's evolution history is obtained through the numerical integration
%% of a standard 1D equation of viscous diffusion but accounting for the
%% collapse of a molecular cloud core \citep{HuGu05}. Unless otherwise
%% mentionned, our fiducial model for the disk evolution is calculated
%% using a turbulent viscosity $\nu=\alpha c_{\rm s} H$ where $c_{\rm s}$
%% is the local sound speed, $H$ is the pressure scale height and
%% $\alpha$ is taken equal to $0.01$.

For simplicity, we consider that our protoplanetary disk is made of 2
components: a dominant one that includes hydrogen, helium and neon
(species that remain gaseous at all temperatures), of abundance
$x_{\rm H}\sim 1$; and a minor one (e.g. argon, krypton or xenon) of
abundance $x_{\rm Z}\equiv x \ll 1$ that is carried with the grains to
the midplane of the disk and radially inward.

\subsection{Enrichment by an inward radial migration of grains?}

The inward radial migration of grains of $\sim$cm size has been
postulated to be a key ingredient in the problem of the formation of
planetesimals \citep[e.g.][]{2004ApJ...614..490C}. Since these grains
should have captured noble gases in the cold outer regions of the
disk, this could potentially also lead to an enrichment of the inner
disk in noble gases. We find however that this scenario is unlikely to
yield noble gases abundances that are compatible with the measured
values in Jupiter. This is due to two reasons: 
\begin{enumerate}
\item Given that we expect a fast mixing of the inner regions, the
  local enrichment of a given species should be proportional to the
  disk mass inside its vaporization radius. Now, this factor is very
  dependent on the vaporization temperature: Compared to a species
  with a 20\,K vaporization temperature, species that would be
  released at 30, 40 or 50\,K should be more enriched by factors
  $\sim$3, 5 and 10, respectively. Given the relatively uniform
  enrichment of Jupiter's atmosphere in Ar, Kr, Xe, and given that Ar
  is much more volatile than Xe, it is difficult to imagine that this
  scenario can work.
\item An efficient transport of noble gases into the disk's inner
  region also implies a rapid loss by accretion onto the star, and a
  global {\it depletion} of these elements. Solving the problem then
  requires a careful balance between inward transport and loss of the
  depleted outer disk (which otherwise will be accreted onto the giant
  planets).
\end{enumerate}

Preliminary mass balance calculations indicate that the inward
transport of noble gases had to be very limited, and is unlikely to
have had an important role: we choose to neglect it in what
follows. We will assume that it may have occurred early on, but led to
the loss of only a minor fraction of the noble gases in the disk.

\subsection{Evaporation from the disk}

\subsubsection{Main characteristics}

It is relatively natural to invoke evaporation of disk material in
order to explain the loss of light hydrogen and helium while retaining
heavier elements.  The fact that disks evaporate has been observed in
the relatively extreme situations of {\it proplyds} under the intense
irradiation of O-B stars, but is also commonly invoked to account for
the fast disappearance of disks around young stars. 

The temperature in the disk atmosphere controls the evaporation, and
defines the critical radius, when the sound speed is equal to the
orbital velocity \citep[e.g.][]{1993Icar..106...92S}:
\begin{equation}
R_g= {G\mstar \mu \over k T},
\label{eq:Rg}
\end{equation}
where $G$ and $k$ are the gravitational and Boltzmann constants,
$\mstar$ is the mass of the central star, and $\mu$ the mean molecular
mass. For a solar-type star and atomic hydrogen, $R_g$ is of order of
10, 100 and 300\,AU, for atmospheric temperatures of $10000$, $1000$
and $350$\,K respectively.

\begin{table*}
\caption{Average characteristics of the evaporation flow estimated at
  the disk's mid-life}
\label{tab:evap}
\begin{tabular}{crrrrrrrrr}\hline\hline
  & & Distance & $n_{\rm midplane}$ & $n_{\rm esc}$ & $T_{\rm esc}$ & $m_{\rm crit}/m_{\rm H}$ &
  $\lambda_{\rm esc}/H_P$ & $\dot{\Sigma}_{\rm esc}$ & $\dot{M}_{\rm esc}$ \\
  & &  [AU] & [cm$^{-3}$] & [cm$^{-3}$] & [K] & [ ] & [ ] & $\rm
 [g\,cm^{-2}\,s^{-1}]$ & $\rm [M_\odot\,a^{-1}]$ \\ \hline
EUV & $\Phi=10^{41}\rm\,s^{-1}$ & $\sim 10$ & $\sim 10^{11}$ & $\sim 10^4$ & $10000$ & $\sim 3000$ & $\sim 10^{-2}$ & $\sim 10^{-14}$ & $\sim 4\times 10^{-10}$ \\
 & $\Phi=10^{43}\rm\,s^{-1}$ &  &  &  & & $\sim 3\times 10^{4}$ & $\sim 10^{-3}$ & $\sim 10^{-13}$ & $\sim 4\times 10^{-9}$ \\ \hline
FUV & $T_{\rm esc}=100$\,K & outer disk & $\sim 10^{10}$ & $\sim 10^5$ & 100 & $\sim 4\times 10^{4}$ & $\sim 3\times 10^{-3}$ & $\sim 10^{-14}$ & $\sim 3\times 10^{-8}$\\
 / & $T_{\rm esc}=200$\,K &            &           &        & 200 & $\sim 3\times 10^{5}$ & $\sim 4\times 10^{-4}$ & $\sim 6\times 10^{-14}$ & $\sim 7\times 10^{-8}$\\
thermal & $T_{\rm esc}=600$\,K &            &           &        & 600 & $\sim 10^{6}$ & $\sim 5\times 10^{-5}$ & $\sim 3\times 10^{-13}$ & $\sim 2\times 10^{-7}$\\
%FUV & $\alpha=3\times 10^{-3}$ & outer disk & $10^{10}$ & $10^5$ & 100 & $\sim 9\times 10^{5}$ & $\sim 5\times 10^{-3}$ & $\sim 10^{-14}$ & $\sim 2\times 10^{-8}$\\
% & \hfill &            &           &        & 200 & $\sim 6\times 10^{4}$ & $\sim 2\times 10^{-3}$ & $\sim 5\times 10^{-14}$ & $\sim 2\times 10^{-8}$\\
% & \hfill &            &           &        & 600 & $\sim 5\times 10^{5}$ & $\sim 10^{-4}$ & $\sim 3\times 10^{-13}$ & $\sim 8\times 10^{-8}$\\
\hline
\end{tabular}
\end{table*}

It is generally thought that only disks that are supercritical
(i.e. $r>R_g$ somewhere in the disk) evaporate
\citep{1993Icar..106...92S, 2000prpl.conf..401H, 2001MNRAS.328..485C},
but \citet{2004ApJ...611..360A} show that an efficient evaporation is
also possible in subcritical disks. In all cases, this requires that a
disk atmosphere is superheated significantly compared to the midplane.
High temperatures ($\sim 10^4$\,K) in the inner disk ($\sim 10$\,AU)
can be due to extreme UV radiation from the central star
\citep{1993Icar..106...92S}; Moderate temperatures ($\sim 100-600\,$K)
in the outer disk can be maintained by the far UV ambient radiation
from a stellar cluster \citep{2004ApJ...611..360A} or by disequibrium
heating of the gas in the presence of some dust particles
\citep{2004A&A...428..511J, 2004ApJ...615..991K}. (Far more violent
evaporation effects can occur near massive stars, but for simplicity,
we choose not to discuss this possibility).

Table~\ref{tab:evap} depicts the global characteristics of the
evaporation flow. A crucial point is that the evaporation is {\it
hydrodynamical}, i.e. it occurs at levels where the mean free path
$\lambda_{\rm esc}$ is significantly smaller than the pressure scale
height $H_P$. Because of the very low gravity, the evaporation also
occurs without any separation of the chemical elements, as shown by
the high value of the critical mass \citep{1987Icar...69..532H}:
\begin{equation}
{m_{\rm crit}/m_{\rm H}}=1+{kT \dot\Sigma_{\rm e}\over b g X_{\rm H}
  m_{\rm H}^2},
\end{equation}
where $b$ is a diffusion coefficient (we use $b\approx 10^{20}\,\rm
cm^{-1}\,s^{-1}$), $g$ is the gravity. 

In all cases, the levels at which evaporation takes place correspond
to relatively high altitudes in the disk, much higher than the
estimated thickness of the dust subdisk \citep{1995Icar..114..237D,
2005A&A...434..971D, 2005ApJ...625..414T, 2005ApJ...623L.149T}.  This
implies that species that condense onto grains and are transported
towards the midplane will generally not be lost from the
disk. Furthermore, because a negative vertical temperature gradient
prevails \citep[e.g.][]{1991ApJ...383..814M, 1997ApJ...490..368C},
even species that are captured by grains at low temperatures but later
vaporize when subject to higher temperatures will not be transported
by convection back up to the levels where they would hydrodynamically
escape. This implies that the disk can become progressively enriched
in species that condense onto grains at low temperatures down to $\sim
15-30$\,K. 

\subsubsection{The model}

Let us consider the evolution of a viscous accretion disk of surface
density $\Sigma(r,t)$, subject to an evaporation rate
$\dot{\Sigma}_{\rm esc}$. Its evolution is governed by the following
equation \citep[e.g.][]{2001MNRAS.328..485C, HuGu05}:
\begin{equation}
{\partial \Sigma\over \partial t}={3\over r}{\partial\over \partial r}
\left[r^{1/2}{\partial\over \partial r}\left(\nu\Sigma r^{1/2}\right)\right]
-\dot{\Sigma}_{\rm esc},
\end{equation}
and the escape rate is limited to a layer of temperature $T_\esc$,
mean molecular mass $m_\esc$, corresponding sound speed $c_\esc$, and
number density $n_\esc$:
\begin{equation}
\dot{\Sigma}_{\rm esc}=m_\esc  c_\esc n_\esc.
\label{eq:Sigma_esc} 
\end{equation}

Let us now consider that a second minor component of surface density
$c\Sigma$ is present (with $c\ll 1$) and obeys the same diffusion
equation as for $\Sigma$, with the exception that for the reasons
discussed previously, its escape rate is very small. (Obviously, this
is a simplification because one should include backreactions between
the 2 species). Within this framework, the evolution of the
concentration $c$, or equivalently of the enrichment ${\cal E}\equiv
c/c_\odot$ (where $c_\odot$ is the protosolar concentration of that
element) can be shown to obey the following equations:
\begin{equation}
{\partial {\cal E}\over \partial t}=3\nu
\left[\left(2{\partial\ln(\nu\Sigma)\over \partial\ln r}+{3\over 2}\right)
{1\over r}{\partial{\cal E}\over \partial r} +
{\partial^2{\cal E}\over\partial r^2}\right] + \dot{\cal E}_{\rm esc}
\label{eq:E}
\end{equation}

\begin{equation}
\dot{\cal E}_{\rm esc}=\left(1-{{\cal E}\over {\cal E}_{\rm max}}\right)
{\cal E}{\dot\Sigma_{\rm esc}\over \Sigma}
\end{equation}

Here, we assume that the enrichment can be at most ${\cal E}_{\rm
max}$, after which both the major and the minor species escape. This
maximal enrichment can be estimated as follows: (i) If the evaporation
takes place inside the noble gases vaporization radius, the
hydrogen-helium atmosphere of the disk is progressively eroded until
the midplane conditions are reached. This implies values of ${\cal
E}_{\rm max}$ of the same order as the midplane dust enrichment
estimated at the vaporization radius. Depending on properties of the
dust and turbulence, it could be either large ($\wig{>}10$)
\citep{2005A&A...434..971D, 2005ApJ...625..414T} or small ($\wig{<}2$)
\citep{1995Icar..114..237D}. (ii) For an evaporation of the outer
disk, the condensation of noble gases onto grains implies that the
scale height of the condensing species $h$ is maintained at a smaller
value than that of the gas, $H$. Because the vertical density profiles
are gaussian, the maximum enrichment can be shown to be of order
${\cal E}_{\rm max}\sim H/h\,{\rm erfc}(z_{\rm esc}/H_P)/{\rm
erfc}(z_{\rm esc}/h)$, where $z_{\rm esc}$ is the altitude at which
escape takes place. The fact that $z_{\rm esc}>H$ in most cases
garanties that ${\cal E}_{\rm max}$ is large (i.e. $\gg 10$), even for
values of $H/h$ close to unity.  Globally, our model rests on the
likely assumption that ${\cal E}_{\rm max}$ is large, and we use a
fiducial value ${\cal E}_{\rm max}=10$.

Note that equation~(\ref{eq:E}) contains both a diffusion term and an
advection term. The latter accounts for the fact that both near the
inner and outer boundaries, matter is essentially advected as it is
either accreted by the star or lost from the system,
respectively. Basically, the problem to be studied is one in which the
disk becomes progressively enriched in heavy elements as it loses
hydrogen and helium, but at a rate that is limited by the diffusion of
these heavy elements in the disk, with a timescale $\sim R^2/\nu$.

We further assume that the accretion of the envelopes of Jupiter and
Saturn is limited by the viscous diffusion in the disk so that the
planets' growth is \citep[see][]{2003ApJ...596L..99L}:
\begin{equation}
\dot{M}_{\rm planet}(t)\equiv e_{\rm planet} \dot{M}_{\rm disk}(r_{\rm
planet},t),
\label{eq:Mplanet}
\end{equation}
where $\dot{M}_{\rm disk}(r_{\rm planet},t)$ is the mass flux in the
disk that crosses the annulus of radius $r_{\rm planet}$ at time
$t$. The parameter $e_{\rm planet}$ is the ratio of the flux accreted
by the planet to the total flux that visously diffuses inward. It is
mostly unknown, and certainly very dependent on the mass of the planet
and its cooling. Based on numerical simulations for Jupiter, we choose
$e_{\rm planet}\sim 0.3$ (F. Masset, pers. communication).

\subsubsection{Inner disk evaporation}

We first follow the approach of \citet{2001MNRAS.328..485C} by
assuming that the young T-Tauri Sun emits extreme UV photons at a rate
$\Phi$, and expecting that $\Phi\approx 10^{41}\,\rm s^{-1}$. In that
case, given that the photons heat the upper layers of the disk to
$\sim 10,000K$, the evaporation mostly takes place at a critical
radius $R_g\sim 10$\,AU \citep{1993Icar..106...92S}. The escape
flux is defined by eq.~(\ref{eq:Sigma_esc}) with $T_\esc\approx
10^4\,$K and by 
\begin{equation}
\left\{ \begin{array}{ll}
n_\esc=n_0 \Phi_{41}^{1/2} R_{g14}^{-3/2} \left(r\over
R_g\right)^{-5/2},\  & \mbox{for }r\ge R_g \\
n_\esc=0 \  & \mbox{otherwise}
\end{array}\right.
\end{equation}
where $n_0=5.7\times 10^4\,{\rm cm}^{-3}$, $\Phi_{41}\equiv
\Phi/10^{41}\rm\,s^{-1}$ and $R_{g14}\equiv R_g/10^{14}\,$cm. The fact
that viscous diffusion is much faster close to the Sun leads to an
inside-out removal of disk material \citep{2001MNRAS.328..485C}.

\begin{figure}
\hspace*{-2.5cm}\resizebox{13cm}{!}{\includegraphics[angle=0]{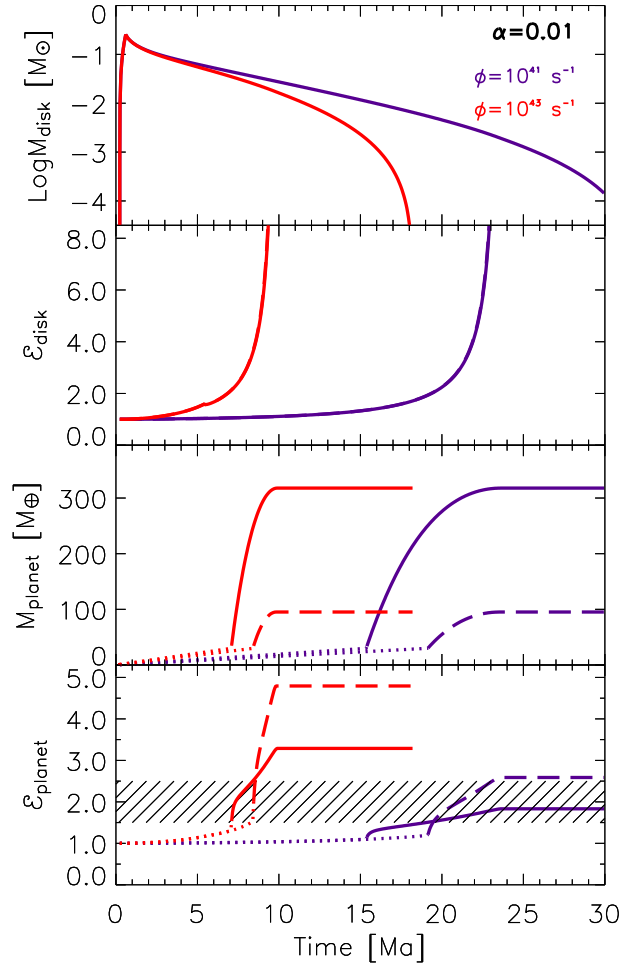}}
\vspace*{-1.2cm}
\caption{Time evolution of the mass of the protosolar disk $M_{\rm
  disk}$, of its local (5 AU) enrichment $\cal{E}_{\rm disk}$, of the
  masses $M_{\rm planet}$ and enrichments $\cal{E}_{\rm planet}$ of
  Jupiter (plain) and Saturn (dashed). The dotted portions of the
  curves correspond to the assumed (slow) growth phase of the planets'
  cores. The calculations are made for a EUV evaporation flux of
  $10^{41}$ (blue) and $10^{43}\,\rm s^{-1}$ (red), respectively. The
  dashed area in the bottom panel indicates the enrichments in Ar, Kr
  and Xe inferred for Jupiter from the Galileo probe data.}
\label{fig:enrich_fig1}
\end{figure}

Figure~\ref{fig:enrich_fig1} shows the resulting evolution of the disk
mass and disk enrichment, the growth of Jupiter and Saturn as
derived from the disk properties and eq.~(\ref{eq:Mplanet}) and the
consequent enrichment of their envelopes, assuming that they are fully
mixed by convection. The effect of evaporation becomes apparent only
when mass of the disk decreases to a point when the mass flux in
the disk becomes comparable to the evaporation mass flux. Because
the evaporation proceeds from inside out, inner regions ($\wig{<}
10$\,AU) are depleted rapidly and the corresponding local enrichment
(2nd panel in fig.~\ref{fig:enrich_fig1}) can become very
large. When this occurs, the mass flux accross these radii has become
tiny, so that Jupiter's enrichment is always found to be moderate and
close to the observed values.

We find that Saturn should be about 50\% more enriched according to
this scenario. However, this crucially depends on the assumption that
Saturn and Jupiter stole the same constant part of the disk mass flux
to build their envelopes (i.e. $e_{\rm Jup}=e_{\rm Sat}=0.3$). If
Saturn has been less efficient at that, then it may be as enriched
in noble gases as Jupiter.

\subsubsection{Outer disk evaporation}

There are two problems with the EUV evaporation scenario: (i) The
timescale for the loss of the circumstellar disk is uncomfortably long
compared to observations, even in the extreme case of
$\Phi=10^{43}\rm\,s^{-1}$; (ii) It requires a high, constant value of
EUV photons production from an unidentified mechanism not powered by
the accretion flow \citep{2003ApJ...582..893M}. 

\citet{2004ApJ...611..360A} hence proposed a scenario based on a
moderate heating of the disks atmosphere by ambient FUV radiation and
a subcritical evaporation. We use a simplified version of their 
evaporation rates (see their appendix) to obtain an evaporation that
is controlled by the temperature of the escaping disk atmosphere:
\begin{equation}
\left\{ \begin{array}{lr}
n_\esc={\sigma_{\rm FUV}^{-1}\over r},\quad & \mbox{for }r\ge R_g \\
n_\esc={\sigma_{\rm FUV}^{-1}\over r}\left(R_g\over r\right)^{1/2}
e^{-(R_g/r-1)/2} \quad & \mbox{otherwise,} 
\end{array}\right.
\end{equation}
where $\sigma_{\rm FUV}\approx 10^{-21}\rm\,cm^{2}$ is the FUV dust
cross section. These escape rates correspond to the evaporation from
the disk surface. The mass lost in the {\it radial} direction is to be
considered and is even dominant for subcritical disks, and we hence
remove the mass to the last radial layer $R_{\rm d}$ of the disk:
\begin{equation}
\left\{ \begin{array}{lr}
\dot M_{\rm rad}=m_\esc  c_\esc \sigma_{\rm FUV}^{-1} R_{\rm
  d},\qquad\qquad & \mbox{for }R_{\rm d}\ge R_g \\ 
\multicolumn{2}{l}{\dot M_{\rm rad}=m_\esc  c_\esc \sigma_{\rm FUV}^{-1} R_{\rm
  d} \left(R_g\over R_{\rm d}\right)^2 e^{-(R_g/R_{\rm d}-1)/2}} \\
 \  & \mbox{otherwise} 
\end{array}\right.
\end{equation}

\begin{figure}
\hspace*{-2.5cm}\resizebox{13cm}{!}{\includegraphics[angle=0]{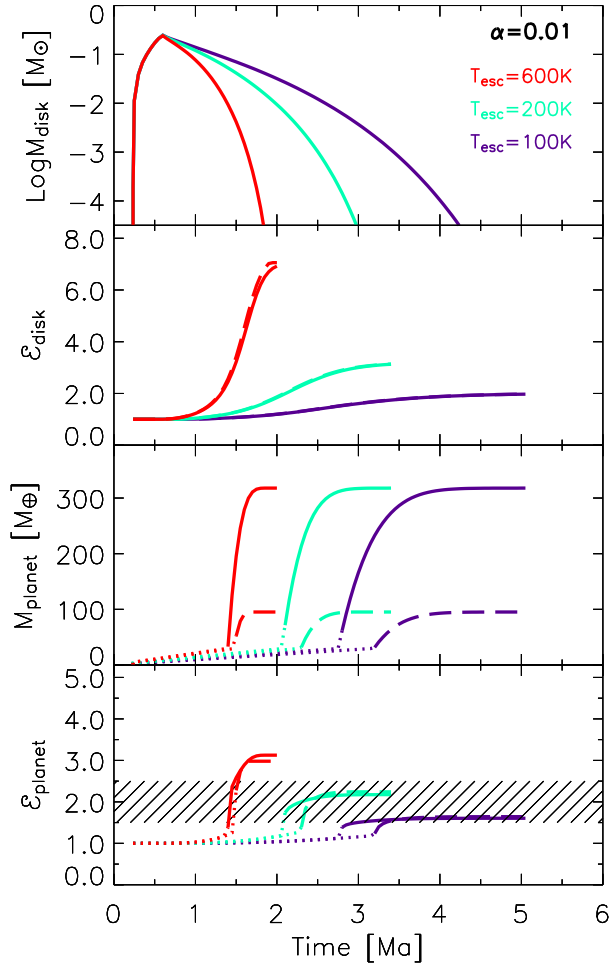}}
\vspace*{-1.2cm}
\caption{Same as fig.~\ref{fig:enrich_fig1}, but in the case of an
  evaporation of the outer disk for three temperatures of the escaping
  disk atmosphere: 100, 200 and 600\,K.}
\label{fig:enrich_fig2}
\end{figure}

Figure~\ref{fig:enrich_fig2} shows the evolutions of the
disk and planets resulting from the combination of early supercritical
and later subcritical evaporation, as the disk progressively
shrinks. As radial evaporation does not lead to an increase in
$\cal{E}$, the enrichment is always self-limited, and mostly
controlled by the value of $T_\esc$ (i.e. by the global ratio of the
vertical to radial evaporation). Once again, we find that for a quite
wide range of values of the different parameters, the resulting
enrichment for Jupiter agrees with the measured enrichment in noble
gases. In this scenario, because the enrichment in the disk is very
progressive, it is expected that Jupiter and Saturn should have about
the same enrichment in noble gases. 

Even though the details of the photoevaporation have been greatly
simplified, we stress that this model does provide an explanation for
the Ma timescales for the removal of circumstellar disks, with viscous
diffusion playing a key role in spreading the disk outward to regions
where it can evaporate more efficiently.

\subsubsection{Sensitivity to input parameters}

We have shown that different scenarii of the evaporation of disks lead
to similar results in terms of final enrichments of the giant
planets. We briefly describe how the most important parameters affect
the results:

{\it Viscosity} ($\alpha$): It affects the global evolution
timescale (proportional to $\alpha$ in the outer-disk evaporation
scenario), but very marginally the final values of the enrichment. 

{\it Initial angular momentum} ($\omega_{\rm cd}$): The maximum disk
mass is directly related to the amount of angular momentum of the
molecular cloud core from which the disk is born. We find that except
for low values of this parameter, the final ${\cal E}_{\rm planet}$
remains quite similar. 

{\it Maximal enrichment} ($\cal{E}_{\rm max}$): Our model rests on
the assumption that the local enrichment in the disk can be relatively
large; the conclusions of the article remain qualitatively valid and
quantitatively similar as long as ${\cal E}_{\rm max}\wig{>} 5$. For
smaller values we expect the final planet enrichment to be smaller
than calculated here and incompatible with the observations. 

{\it Planet growth factor} ($e_{\rm planet}$): This parameter governs
the growth of our model planets. Because the enrichment in the disk
rapidly increases, larger values of this factor yields larger
$\cal{E}_{\rm planet}$. However, the dependance is only moderate
because of the rapid decline in the disk (and planet) accretion rate.
We tested that changes are smaller than the error bars on the
abundances of noble gases for values of $e$ between $0.1$ and $1$.

\section{Conclusions}

We have shown that the observed enrichment of Jupiter in noble gases
can be explained by a progressive capture of the planet's envelope in
sync with the evaporation of the protosolar disk. We infer that the
envelope capture phase started relatively late, at about half of the
disk's lifetime, probably because of a slow or delayed growth of the
protoplanetary core. In our scenario, the final masses of Jupiter and
Saturn result naturally from a competition between limited viscous
accretion and disk evaporation. We note that a welcome consequence of
the late start scenario is a significant suppression of the inward
migration when compared to models in which giant planets form early,
in locally massive disks (such as the minimum mass solar nebula or
more). Provided disk atmospheres can be heated to temperatures of
100\,K or more, the disk model that we propose further explains the
disappearance of disks in Ma timescales, as observed.

We predict that the enrichments in Ar, Kr, Xe in Saturn should be
similar to that in Jupiter. It should be larger in Uranus and Neptune,
due to a late capture of more enriched gas from the evaporating disk,
but also to a pollution of unknown magnitude from their ice cores,
significantly more massive than their gaseous envelopes. No isotopic
fractionation is expected to arise from the disk evaporation
process. However, clues on the ambiant mean temperature of the
protosolar molecular cloud core may be obtained from the primordial
Ne/Ar, Ne/Kr, and Ne/Xe ratios: if this mean temperature was high
enough (e.g. $20\sim 30$\,K) not to allow neon to condense onto
grains, then neon should have escaped progressively from the disk with
hydrogen and helium, yielding small values of these ratios. In all
cases these ratios should be equal to or smaller than unity.

These results show the importance of noble gases for tracing back
events that occurred in the early Solar System and stress the need for
an accurate determination of the compositions of all giant planets.

\bibliography{jupcomp}

\bsp

\label{lastpage}

\end{document}